\documentclass[12p]{revtex4}
\usepackage{epsfig}
\usepackage{mathrsfs}
\usepackage{amsmath}
\usepackage{amsfonts}
\usepackage[flushleft,small]{caption2}
\usepackage{wrapfig}
\usepackage{CJK}
\usepackage{graphicx}
\usepackage{booktabs}

\def\d{\mathrm{d}}

\def\ben{\begin{equation}}
\def\een{\end{equation}}
\def\bea{\begin{eqnarray}}
\def\eea{\end{eqnarray}}

\begin{document}
\title{Cosmological Dynamics of de Sitter Gravity \footnote{Supported by the Specialized Research Fund for the Doctoral Program of Higher Education (SRFDP) under Grant No 200931271104, Shanghai Municipal Pujiang Foundation under Grant No 10PJ1408100.
$^{**}$Email: kychz@shnu.edu.cn}}
\author{ Xi-chen Ao, Xin-zhou Li, Ping Xi}
\affiliation{Shanghai United Center for Astrophysics (SUCA),  Shanghai Normal University, 100 Guilin Road, Shanghai 200234 }

\begin{abstract}
A new cosmological model based on the de Sitter gravity
is investigated by dynamical analysis and numerical discussions.
Via some transformations, the evolution equations of this model can form an
autonomous system with 8 physical critical points. Among these critical
points there exist one positive attractor and one negative attractor.
The positive attractor describes the asymptotic behavior of late-time
universe, which indicates that the universe will enter the exponential
expansion phase, finally. Some numerical calculations are also
carried out, which convince us of this conclusion derived from the
dynamical analysis.\\[0.4cm]

\end{abstract}
\maketitle

In the last decades, some new cosmological observations such as SNeIa, CMBR and large scale structure
all indicate that our universe is accelerating expanding and there exists a new mystical energy component in our universe, dubbed dark energy.\cite{dark energy} Many heuristic models have been proposed to explain the nature of this new component, which account for almost 74\% of the energy density of our universe. Some of them have physical foundations,\cite{theoretical} some are just phenomenological.\cite{phenomena} However, no one is flawless. All these models have their own problems, such as cosmological constant problem and fine tuning problem. Recently, a new kind of model, called torsion cosmology,
has drawn researchers' attention, which is basically based on some new gravity theory, the gauge theory of gravity. Among these
models, Poincare gauge theory is the one that has been investigated widely, which is inspired from the Einstein special relativity and the localization of Poincare symmetry.\cite{PGT} Nester {\it et al.}\cite{Nester} applied this new gravity theory to cosmology, and obtained a novel model which is likely to be a new explanation of the accelerating expansion. In that model, dynamic connection mimics the contribution of dark energy. Based on the work of Nester and his colleagues, some dynamics analysis and analytical discussion have been conducted in many papers, from which we can know the fate of the universe more clearly.\cite{Li}

Besides Poincare gauge theory, there is another classical gauge theory of gravity, de Sitter gravity, which can also be the alternative gravity theory for Einstein Gravity. This theory is derived from the de Sitter invariant special relativity and the localization of de Sitter symmetry.\cite{dsg1} In the de Sitter gravity theory, Lorentz connection and tetrad are combined to form a new connection, i.e. dS connection, which is valued in $\mathfrak{s o}$(1,4), rather than Lorentz connection's $\mathfrak{s o}$(1,3); and the gravitational action takes the form of Yang-Mills gauge theory.  Like Poincare gauge theory, the spacetime also has a generic Riemann-Cartan structure, $U_{4}$. From the variational principle one can obtain the gravitational field equation. Analogous to PG theory, de Sitter gravity has also been applied to the cosmology recently to explain the accelerating expansion.\cite{huang}

In this Letter, we transform the cosmological evolution equations of de Sitter gravity model into a set of first-order dynamics equations, which form  an autonomous system. Then, we give some dynamical analysis of this autonomous system and obtain all the critical points. We analyze the critical points' dynamical properties and stabilities, and  find out that there exists a late-time de Sitter attractor. It is concluded that the universe will expand exponentially in the end, as the $\Lambda \mathrm{CMD} $ predicted.

For a homogeneous, isotropic universe, the space-time is described by the FLRW metric:
 \begin{eqnarray}\label{frw}
 ds^2=dt^2-a^2(t)[\frac{dr^2}{1-k r^2}+r^2(d \theta^2+ \sin^2 \theta d
 \phi^2)],
 \end{eqnarray}
 and the isotropic and homogeneous torsion takes the form
 \begin{eqnarray}
{\bf T}^0 &=& 0 \nonumber \\
{\bf T}^1 &=& {T_+}\,  {\vartheta}^0\wedge {\vartheta}^1 +
{T_-}\,  {\vartheta}^2\wedge {\vartheta}^3\nonumber \\
{\bf T}^2 &=&  {T_+}\,  {\vartheta}^0\wedge {\vartheta}^2 -  {T_-}\,  {\vartheta}^1\wedge {\vartheta}^3 \label{torsion} \\
{\bf T}^3 &=& {T_+}\,  {\vartheta}^0\wedge {\vartheta}^3 +  {T_-}\,  {\vartheta}^1\wedge {\vartheta}^2 , \nonumber
 \end{eqnarray}
where $T_+$ is the trace part of the torsion, $\frac 1 3 T^a_{\ 0a}$,
while $T_-$ is the traceless part of the torsion. They are all functions of time $t$, with + and $-$
denoting the even and odd parities, respectively. Here $\vartheta^{0}=\d t,\ \vartheta^{1}=a(t)\d r,\ \vartheta^{2}=a(t)r\d \theta\ \mathrm{and}\ \vartheta^{3}=a(t)\sin\theta \d \phi $.

According to the field equations of de Sitter gravity and Eqs.(\ref{frw}) and (\ref{torsion}), one can easily obtain the new evolution equations of universe,\cite{huang}
 \begin{eqnarray}
\label{el-00}%
&& - \frac {\ddot a^2} {a^2}
 -  \left(\dot T_++ 2\frac{\dot a}{ a} T_+ -2\frac {\ddot a}{a} \right)\dot T_+
 + \frac1 4 \left(\dot T_-+2\frac{\dot a} aT_- \right)\dot T_-
+ T_+^4-\frac {3}{2}
T_+^2 T_-^2+ \frac{1} {16} T_-^4 + \left(5 \frac{\dot a^2} {a^2}\right. \nonumber \\
&& \quad \left. + 2\frac{ k} {a^2}-\frac{3}{R^2}\right) T_+^2-\frac{1}{2}
\left(\frac{5}{2}\frac{\dot a^2} {a^2} + \frac{k}{a^2}  -\frac{3}{R^2}\right) T_-^2 + 2 \frac
{\dot a} a \left(\frac{\ddot a} {a}  - 2 \frac{\dot a^2} {a^2}-2\frac{ k} {a^2}
+\frac{3}{R^2}\right)T_+ - \frac{\dot a} {a} (4  T_+^2 \nonumber\\
&&\quad  - 3  T_-^2)T_+ +\frac{\dot a^2}{a^2}\left( \frac
{\dot a^2}{a^2}  + 2 \frac{k} {a^2}- \frac{2}{R^2}\right) +\frac{k^2}{a^4}  - \frac{2}
{R^2} \frac{k}{a^2} +\frac{ 2}{R^4}=-\frac{16\pi G\rho}{3 R^2},
 \end{eqnarray} %
 \begin{eqnarray} \label{el-11}%
&&\frac{\ddot a^2} {a^2} + \left(\dot T_+ + 2\frac{\dot a} a  T_+ - 2\frac{\ddot a} a
+ \frac{6}{R^2}\right)\dot T_+ -\frac 1 4 \left(\dot T_-
+ 2 \frac {\dot a} a T_-\right)\dot T_- - T_+^4 + \frac 3 2 T_+^2 T_-^2 - \frac1 {16} T_-^4 \nonumber\\
&&\quad+ \frac {\dot a} a(4  T_+^2 - 3 T_-^2)T_+  - \left(5\frac{\dot a^2} {a^2} + 2 \frac k
{a^2}  + \frac3 {R^2}\right) T_+^2+ \frac 1 2 \left(\frac 5 2\frac{\dot a^2} {a^2} +
\frac k {a^2} +  \frac 3 {R^2}\right) T_-^2- 2\frac{\dot a} a \left(\frac{\ddot a } {a}- 2\frac{\dot a^2} {a^2 }\right.\nonumber \\
&&\quad \left. - 2  \frac k {a^2}- \frac6 {R^2}\right)T_+  - \frac 4 {R^2} \frac{\ddot a} a -\frac{\dot a^2} {a^2} \left(\frac{\dot a^2}{a^2} +2\frac k
{a^2}\right )+  \frac2 {R^2}
-\frac{k^2}{a^4}  - \frac2 {R^2}\frac k {a^2} +\frac6 {R^4}
= -\frac{16\pi G p}{R^2},%
 \end{eqnarray}
 \begin{eqnarray}\label{yang1} %
&&\ddot T_-  + 3 \frac{\dot a} a \dot T_- + \left( \frac 1 2 T_-^2 - 6 T_+^2 + 12 \frac {\dot a} a
T_+  +\frac{\ddot a} a - 5\frac{\dot a^2}{a^2}
-  2\frac k {a^2}+  \frac 6 {R^2}\right)  T_-=0,
 \end{eqnarray}%
 \begin{eqnarray}\label{yang2}%
&&  \ddot T_+ + 3 \frac{\dot a} a \dot T_+ -\left( 2  T_+^2  -\frac 3 2 T_-^2 - 6\frac{\dot a} a
T_+ -\frac {\ddot a} a  + 5 \frac {\dot a^2}  {a^2} + 2 \frac k {a^2}- \frac 3 {R^2}\right)
T_+ - \frac 3 2 \frac{\dot a} a T_-^2-\frac{\dddot a} a - \frac{\dot a\ddot a} {a^2} + 2\frac {\dot a^3} {a^3}
 + 2\frac{\dot a} a
\frac k {a^2} =0. %
 \end{eqnarray}%
Equations\,(\ref{el-00}) and (\ref{el-11}) are the 0-0 and 1-1 component of Einstein-like equations, respectively; and Eqs.(\ref{yang1})-(\ref{yang2}) are two independent Yang-like equations, which are derived from the $(r,\theta,\phi)$ and $(t,r,r)$ components. Here we also assume that the spin density is zero.

Also, the energy momentum tenor is conserved by the virtue of Bianchi identities, leading to the continuity equation
 \begin{eqnarray}
\dot{\rho}&=&-\frac{3\dot{a}}{a}(\rho+p). \label{continuity}
 \end{eqnarray}
Equation (\ref{continuity}) can be derived from Eqs.(\ref{el-00})--(\ref{yang2}), which means 4 of Eqs.(\ref{el-00})--(\ref{continuity}) are independent.
These four independent equations with the EOS of matter content comprise a complete system of equations for five variables $a(t),\ T_-(t),\ T_+(t),\ \rho(t)$ and $p(t)$.
By some algebra and differential calculations, we could simplify these five equations to
 \begin{eqnarray}
\dot{H}&=&-2H^{2}-\frac{k}{a^{2}}+\frac{2}{R^{2}}+\frac{4\pi G}{3}(\rho+3p)+\frac{3}{2}\left(\dot{T}_++3H T_{+}-T_{+}^{2}+\frac{T_-^2}{2}\right)+(1+3w)\rho ,\\
\ddot{T}_+&=&-3\left(H +\frac 3 2 T_+\right)\dot{T}_+ -3T_{-}\dot{T}_- -\frac{8\pi G}{3}(\rho+3p)^.
-\frac{3}{2}H T_{-}^{2}+\left[\frac {13} 2 ({T_+}-3 H){T_+}+ 6H^2+\frac{3k}{a^2}\right.\nonumber \\
&&
\left.+\frac{9T_-^2}{4}-\frac 8 {R^2}-\frac{28\pi G}{3}(\rho+3p)\right]T_+ ,\\
\ddot{T}_- &=&-3H\dot{T}_- -\left[-\frac{15}{2}T_{+}^{2}+\frac{33H T_{+}}{2}-6H^{2}-\frac{3k}{a^2}+\frac{8}{R^{2}}+\frac{5}{4}T^{2}_{-}+\frac{3}{2}\dot{T}_+ + \frac{4\pi G}{3}(\rho+3p)\right]T_{-},\\
\dot{\rho}&=&-3H(\rho+p),\\
w&=&\frac{p}{\rho},
 \end{eqnarray}
where $H=\dot{a}/a$ denotes the Hubble parameter.
In order to make these equations dimensionless, we can rescale the variables and parameters as
\begin{eqnarray}
&&t\rightarrow t/l_0;\quad H\rightarrow l_0 H;\quad k\rightarrow l_0^{2}k;\quad R\rightarrow R/l_0;\nonumber \\
&&T_{\pm}\rightarrow l_0 T_{\pm};\quad \rho \rightarrow \frac{4\pi G l_{0}^2}{3 }\rho;\quad p \rightarrow \frac{4\pi G l_{0}^2}{3 }p,\label{transformation}
\eea
where $l_0=c/H_0$ is the Hubble radius.

By some further calculations, we find that if the equation of state is constant, then these equations would turn out to form a six-dimensional autonomous system, which takes the forms
\begin{eqnarray}\label{dia}
\dot{H}&=&-2H^{2}+\frac{2}{R^{2}}-k\left(\frac{\rho}{\rho_0}\right)^{\frac {2}{ 3(1+w)}}+\frac{3}{2}\left(P+3H T_{+}-T_{+}^{2}+\frac{T_-^2}{2}\right)+(1+3w)\rho ,
\label{dit}
\eea
\begin{eqnarray}
\dot P &=&
-3\left(H +\frac 3 2 T_+\right)P -3T_{-}Q-\frac{3}{2}H T_{-}^{2}+ \left[\frac {13} 2 ({T_+}-3 H){T_+}+ 6H^2+3k\left(\frac{\rho}{\rho_0}\right)^{\frac {2}{ 3(1+w)}}
-\frac 8 {R^2}
- 7\rho\right]T_+\nonumber \\&& + 6H(1+w)\rho ,\\[0.2cm]
\dot{T_{+}}&=&P,\\
\dot Q&=&-3H Q -\left[-\frac{15}{2}T_{+}^{2}+\frac{33H T_{+}}{2}-6H^{2}-3k\left(\frac{\rho}{\rho_0}\right)^{\frac{2}{ 3(1+w)}}+\frac{8}{R^{2}}+\frac{5}{4}T^{2}_{-}+\frac{3}{2}P + (1+w)\rho \right]T_{-},\\
\dot{T_{-}}&=&Q,\\
\dot{\rho}&=&-3H(1+w)\rho ,\label{rho}
\end{eqnarray}
where $\rho_0$ is a dimensionless parameter denoting the current energy density.
For such an autonomous system, i.e. Eqs.(\ref{dia})--(\ref{rho}), we can use the qualitative method of ordinary equations with respect to the new set of variables, $(H,\,P,\,Q,\,T_{+},\,T_{-},\,\rho) $.
In order to find the critical points of this system, we should set the left-hand side of Eqs.(\ref{dia})--(\ref{rho}) to zero and solve these algebra equations. Next, we just consider the matter dominant case with spatial flatness. We find that there are nine critical points ($H_c,P_{c},Q_{c},T_{+ c},T_{- c},\rho_{c}$) of this system, as shown in Table 1.
\\[0.2cm]

Furthermore, we analyze the stability of these critical points by means of first-order perturbation. By the Taylor expansion, we could obtain the perturbation equation around the critical points, i.e.
\begin{align}
\delta \boldsymbol{\dot{x}} = A \boldsymbol{x}, &  \quad A = \frac{\partial \boldsymbol{f}}{\partial \boldsymbol {x}}|_{\boldsymbol{x}=\boldsymbol{x}_c},
\end{align}
where $\boldsymbol{x}$ means the six variables of this autonomous system and $\boldsymbol{f}$ denotes the six vector functions on the right-hand side of Eqs.\,(\ref{dia})--(\ref{rho}).
According to the dynamical analysis theory, we could classify these critical points by the coefficient matrix $A$'s eigenvalue.
The classification of these critical points are shown in Table 2. It is easy to find that there are only one positive attractor, i.e. point 1, whose eigenvalues all are negative, and only one negative attractor, i.e. point 2, whose eigenvalues all are positive. Positive attractors attract the solutions nearby it, while negative attractors repel them. The phase lines connecting a positive attractor and a negative attractor are called the heteroclinic lines as shown in Fig.\,1.

\begin{table}[t]
\begin{tabular}{l c c l}\hline\hline
 &Critical points & Eigenvalues  \\[0.1cm]
\hline\\[-0.15cm]
(1)&$(\frac{1}{R},0,0,0,0,0)$&
$-\frac{1}{R},-\frac{1}{R},-\frac{2}{R},-\frac{2}{R},-\frac{3}{R},-\frac{4}{R}$\\[0.12cm]
(2)&$(-\frac{1}{R},0,0,0,0,0)$&
$\frac{1}{R},\frac{1}{R},\frac{2}{R},\frac{2}{R},\frac{3}{R},\frac{4}{R}$ \\[0.12cm]
(3)&$(-\frac{1}{2R},0,0,-\frac{2}{R},0,0)$&
$-\frac{2}{R},\frac{2}{R},\frac{3}{2R},-\frac{5}{2R},\frac{7}{2 R},\frac{4}{R}$\\[0.12cm]
(4)&$(\frac{1}{2R},0,0,\frac{2}{R},0,0)$&
$-\frac{2}{R},\frac{2}{R},\frac{5}{2R},-\frac{3}{2R},-\frac{4}{R},-\frac{7}{2R}$\\[0.12cm]
(5)&$(-\frac{1}{2R},0,0,\frac{1}{2R},0,0)$&
$\frac{1}{2R},\frac{1}{R},-\frac{1}{R},\frac{2}{R},\frac{3}{2R},\frac{5}{2R}$\\[0.12cm]
(6)&$(\frac{1}{2R},0,0,-\frac{1}{2R},0,0)$&
$-\frac{1}{2R},\frac{1}{R},-\frac{1}{R},-\frac{3}{2R},-\frac{5}{2R},-\frac{2}{R}$
\\[0.12cm]
(7)&$(0,0,0,-\frac{\sqrt{3/2}}{R},0,\frac{1}{4R^{2}})$&
$\frac{-1.661}{R},\frac{1.747}{R},\frac{-1.691 -0.447\mathrm{i}}{R},$\\[0.12cm]
  & &$\frac{-1.691 +0.447\mathrm{i}}{R},\frac{3.297}{R},0$\\[0.12cm]
(8)&$(0,0,0,\frac{\sqrt{3/2}}{R},0,\frac{1}{4R^{2}})$&
$\frac{1.661}{R},\frac{-1.747}{R},\frac{1.691 -0.447\mathrm{i}}{R},$\\[0.12cm]
  & &$\frac{1.691 +0.447\mathrm{i}}{R},\frac{-3.297}{R},0$\\[0.12cm]
(9)&$(0,0,0,0,0,\frac{-2}{R^{2}})$ & Not physical  &  \\[0.10cm]
\hline\hline
\end{tabular}
\caption{  The critical points and their corresponding eigenvalues. The point 9 is not physical, for its negative energy density.}
\end{table}

Critical points are actually some exact solutions of a dynamic system, especially the positive attractors which are often the extreme points of the orbits in  phase space. Therefore, they would describe the asymptotic behaviors of solutions. In this model, the positive attractor point 1 would show us the late time evolution of our universe. It is indicated that all quantities tend to zero, except the Hubble constant, which will remain at a fixed value, and therefore the whole universe will approach the exponential expansion, just like the $\Lambda$CDM model.
\begin{eqnarray}
H_\infty=\frac{\dot{a}}{a}=\frac{1}{R} \Rightarrow a(t)\propto \mathrm{exp}\left(\frac{t}{R}\right)
\end{eqnarray}
\begin{table}[t]
\begin{tabular}{l c c l}\hline\hline
 Critical points & Property & Stability \\[0.1cm]
\hline\\[-0.15cm]
 (1)& Positive-attractor & Stable\\[0.12cm]
 (2)& Negative-attractor & Unstable\\[0.12cm]
 (3)& Saddle & Unstable\\[0.12cm]
 (4)& Saddle & Unstable\\[0.12cm]
 (5)& Saddle & Unstable\\[0.12cm]
 (6)& Saddle & Unstable\\[0.12cm]
 (7)& Spiral-saddle&  Unstable\\[0.12cm]
 (8)& Spiral-saddle& Unstable\\[0.12cm]
\hline\hline
\end{tabular}
\caption{  The stability properties of critical points.}
\end{table}
\begin{figure}[t]
\centering
\includegraphics[width=5.5cm,height=4.5cm]{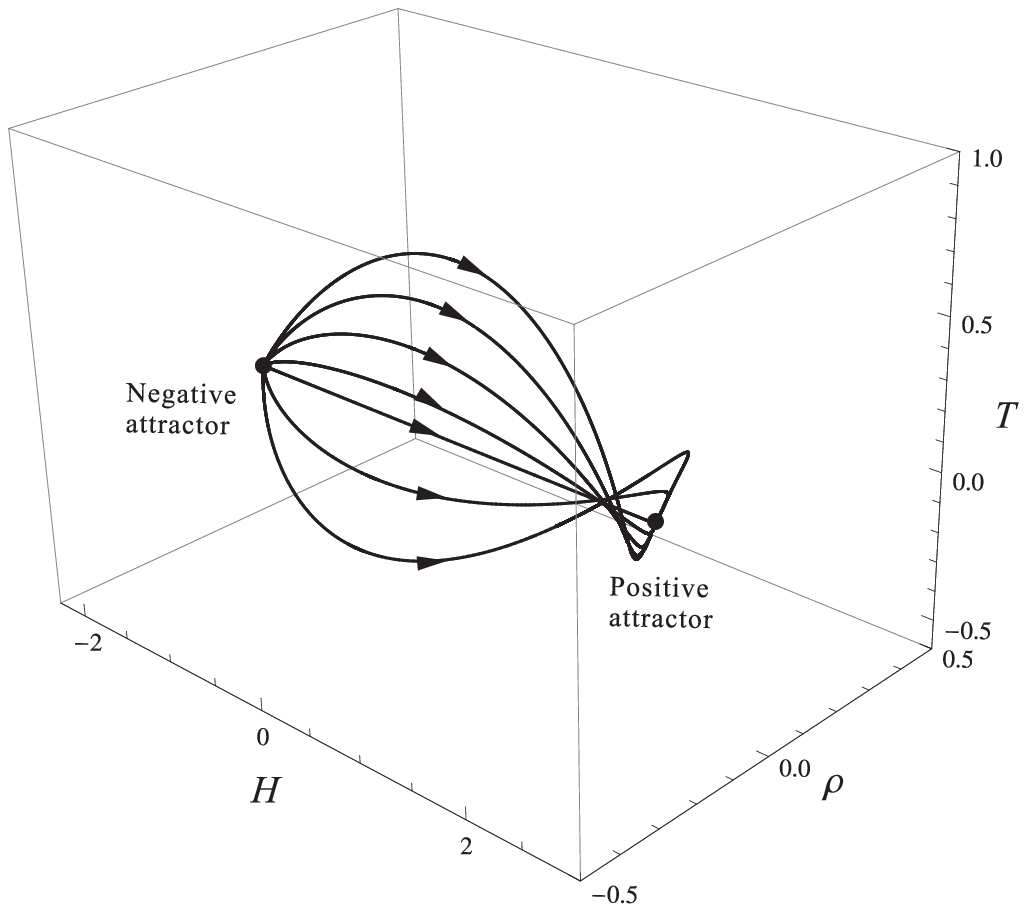}
\caption{ The $(H,T_+,\rho)$ section of the phase diagram with $R=4/3$. The heteroclinic orbits connect the critical points 1 and 2.}
\end{figure}

As we know, the critical points can only describe the local properties. If we want to know the global properties, we have to resort to the numerical calculations.
We solve Eqs.(\ref{dia})--(\ref{rho}) numerically, and show some generic solutions in Fig.\,2. From these numerical results, we could see easily that the late-time positive attractor covers a wide range of initial values, and therefore alleviate the fine-tuning problem. Also for its insensitivity to the initial conditions, this autonomous system has no chaotic features. From this perspective, the cosmology based on de Sitter gravity is quite different from the one based on the PG theory, which suggests that the expansion will asymptotically come to a halt.\cite{Li} However, this discrepancy is only due to the existence of the de Sitter radius $R$.
If we set $R\rightarrow \infty$, the de Sitter gravity would degenerate to the PG theory, and have the same conclusion.
\begin{figure}[t]
\includegraphics[width=7cm,height=4.3cm]{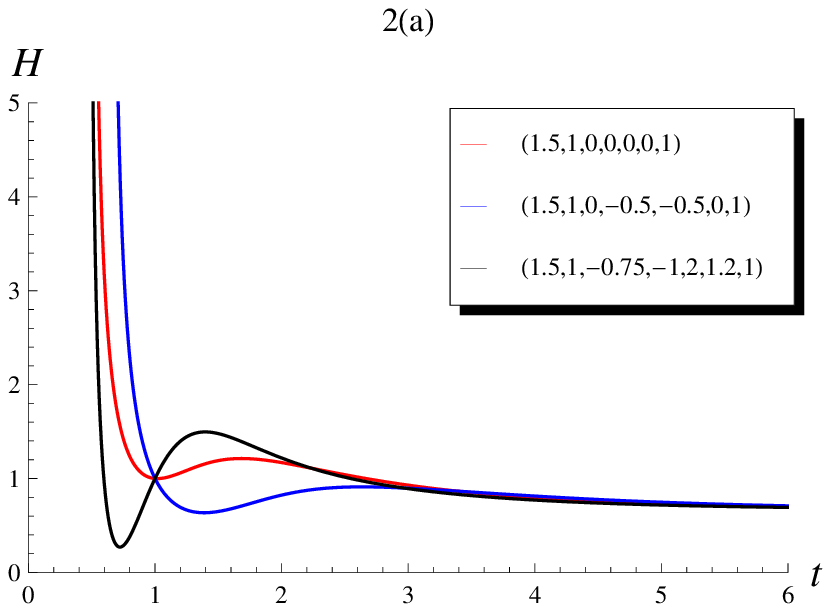}
\includegraphics[width=7cm,height=4.3cm]{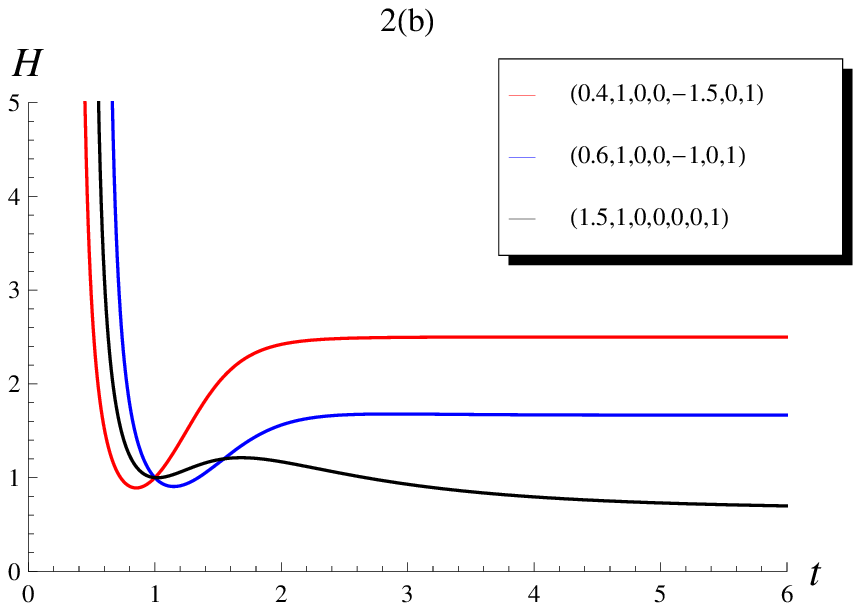}
\caption{ Evolution of Hubble constant $H$ with respect to some initial values and parameter choice $(R,H_0,P_0,Q_0,T_{+0},$ $T_{-0},\rho_0)$. According to the transformations (\ref{transformation}), the unit of time here is the Hubble Time. (a) $R$ is fixed and $T_\pm$ is changed. (b) $R$ is changed.}
\end{figure}
\vspace*{0.1cm}

In summary, we have studied the torsion cosmology based on de Sitter gravity. According to Ref.\cite{huang}, we have rewritten the evolution equations as a set of first-order dimensionless equations which form a six-dimensional autonomous system. We find out that among all the eight physical critical points, there are one positive attractor and one negative attractor. The positive attractor implies that the universe will expand exponentially in the end and all other physical quantities will turn out to vanish. Also we present some numerical analysis of this model, and find out that the late-time evolution is not sensitive to the initial values and parameter and, for a large range of parameter choice, the dynamical system would approach to the positive attractor.  Therefore, in this sense, the de Sitter gravity model looks more like the $\Lambda\mathrm{CDM}$ model,\cite{theoretical} rather than the PG theory.\cite{Li}

If we want to know deeper on whether this model can explain the accelerating expansion, we have to settle the initial values and parameter choice. It requires us to do
some further analytical and numerical calculations and examine them with the current observations, such as SNeIa and CMB etc. These issues will be considered in the upcoming papers.

\end{document}